# IR-UWB Channel Capacity for Analog and Mostly Digital Implementation

Aubin Lecointre[1], Daniela Dragomirescu[1,2], and Robert Plana[1,2].
1 - LAAS-CNRS; Université de Toulouse;7, Av du Colonel Roche
F-31077 Toulouse, France
2-Université de Toulouse ; INSA, UPS

*Abstract*—The impact of the type of implementation is considered on the IR-UWB channel capacity. This study is lead for analog and mostly digital implementation. Key parameters and theirs impacts on the channel capacity are exposed in each case: data converters for mostly digital implementations and pulse generators capabilities for analog implementations. These two implementations are compared from a data rate point of view. Their behaviors regarding an increase of the operating frequency are also studied.

*Index Terms*—Channel capacity, IR-UWB, mostly digital radio, implementation considerations, A/D converters.

## I. INTRODUCTION

THIS paper proposes a study of the IR-UWB (Impulse Radio Ultra WideBand) channel capacity, by using a new expression obtained from Shannon capacity [1], which takes in consideration the kind of implementation. IR-UWB could be designed in a mostly digital radio way or in an analog way [2]. This study exposes architecture key points and their importance from a high data rate point of view. By proposing to merge the IR-UWB channel capacity study and implementation considerations, we are able to specify the dimensioning element for each kind of architectures. Achievable data rate values, for analog and mostly digital implementation, are also obtained thanks to the IR-UWB channel capacity. In this article only binary modulations will be considered, in order to emphasize the IR-UWB simplicity behaviour.

This paper is laid out as follow: in Section II, we exhibit the IR-UWB channel capacity for a mostly digital implementation, while Section III is dedicated to analogue implementation and section IV is devoted to conclusions.

This work was supported by the ANR, the French research national agency. Authors thank ANR for supporting this work, in the framework of the RadioSOC project (n°JC05-60832). A. Lecointre thanks DGA, "Délégation Générale pour l'Armement", for PhD funding.

## II. MOSTLY DIGITAL IMPLEMENTATION LIMITATIONS ON IR-UWB CHANNEL CAPACITY

If we consider a mostly digital implementation for IR-UWB as described in fig. 1, the IR-UWB channel capacity is now dimensioned by the channel delay spread and also by key points of this architecture. They are analog-to-digital converters (ADC), digital-to-analog converters (DAC) and the digital part dedicated for the digital signal processing: FPGA (Field Programmable Gate Array) or ASIC (Application Specific Integrated Circuit).

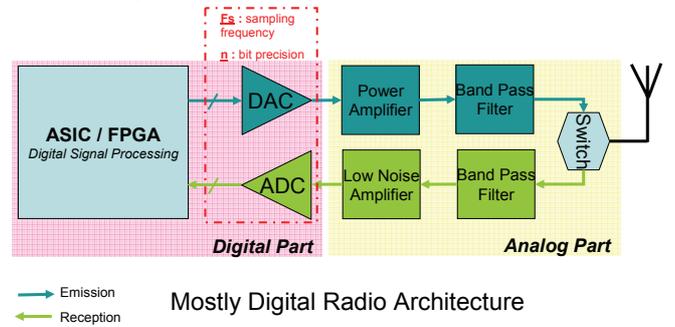

Fig. 1 – The mostly digital radio implementation for an IR-UWB radio. Illustration of ADC/DAC as a key compenent.

Performances of this part will dimension performances of the transceiver and also the achievable data rate. That's why we propose to determine the IR-UWB channel capacity (1), in the binary modulation case, for a mostly digital radio in function of the sampling frequency ($F_s$) used in ADC/DAC.

$$C_{IR-UWB\,MOSTLY-DIGITAL}(bits/s) = \frac{1}{\frac{n_{sampling}}{F_s} + d_{RMS}} \quad (1)$$

With $n_{sampling}$ the sampling factor, i.e. the ratio between the sampling frequency and the analog signal maximum frequency (the inverse of the IR-UWB pulse duration); $F_s$ the sampling frequency of data converters, and $d_{RMS}$ the RMS channel delay spread.

The digital circuit frequency is not a dimensional element, since thanks to techniques such as time interleaved ADC, the digital signal processing is done at a lower frequency but with a parallelization of the processing and a retiming algorithm at the output of data converters [3] [4]. As a result only sampling frequency of the data converter has to take into consideration in the evaluation of the channel capacity, since it's the most dimensioning parameter.

The binary modulation framework is reinforced, in mostly digital implementation, by the flash ADC capability. Flash converters, for power consumption and surface reasons can achieve a high sampling frequency or a high bit precision but not both [5]. Due to the IR-UWB bandwidth we are forced to consider high speed data converters and thus binary modulations for respecting flash converters capabilities.

An analysis of the IR-UWB channel capacity regarding the delay spread, the sampling frequency and the sampling factor is exposed in figure 2.

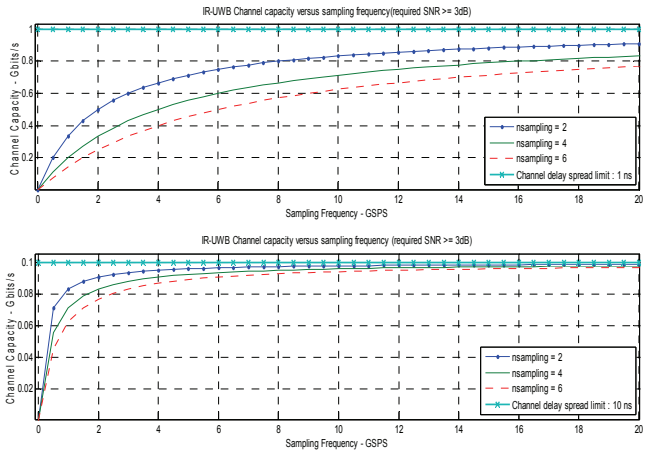

Fig. 2 – IR-UWB channel capacity for mostly digital radio implementation for three distinct channel delay spreads. Impact of sampling frequency and sampling factor is exposed.

For mostly digital implementation, the delay spread remains the most important limitation. It defines an asymptote at $1/d_{RMS}$. The higher the sampling frequency is, the higher the achievable data rate is. Low delay spread channels require higher sampling frequencies, for yielding the channel capacity than high delay spread channels (figure 3)

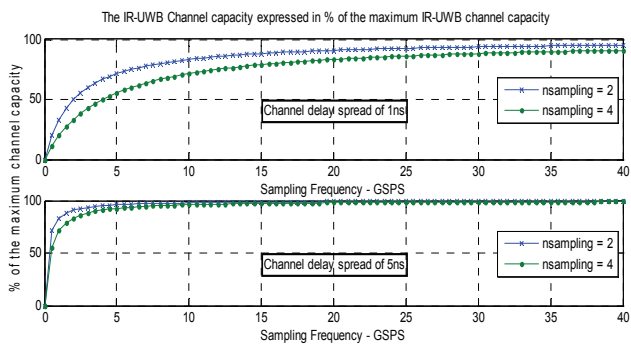

Fig. 3 – IR-UWB channel capacity for mostly digital radio implementation expresses in percent of the maximum capacity.

Figure 3 shows that the achievable data rate difference for two successive low sampling frequencies is larger than for two successive higher sampling frequencies. This is due to the presence of the delay spread asymptote. An increase of the sampling frequency for low sampling frequencies is efficient regarding the data rate.

Concerning the sampling factor, the higher it is, the lower the channel capacity is. An increase of the sampling factor implies a decrease of the channel capacity, but it permits to achieve better performances regarding the bit error rate (BER) and, for example, the synchronization precision [6]. Thus, there is a data rate versus performance trade-off. Fig. 2 exposes also that the gain of low sampling factors, for achieving high data rates, decrease with an increase of the sampling frequency. This is due to the delay spread asymptote behaviour of the IR-UWB channel capacity.

Achievable data rate values for IR-UWB mostly digital implementations can be determined by using the state of the art regarding data converter performances [7] and the sampling factor. This sampling factor has generally a value of four [6], this value is adapted for achieving a good balance regarding correlation and synchronization performances while minimizing the power consumption. For obtaining realistic IR-UWB channel delay spreads, the IEEE 802.15.4a channel model is use.

Figure 4 and Table I present these achievable data rate values if we consider state of the art components and realistic IR-UWB channels.

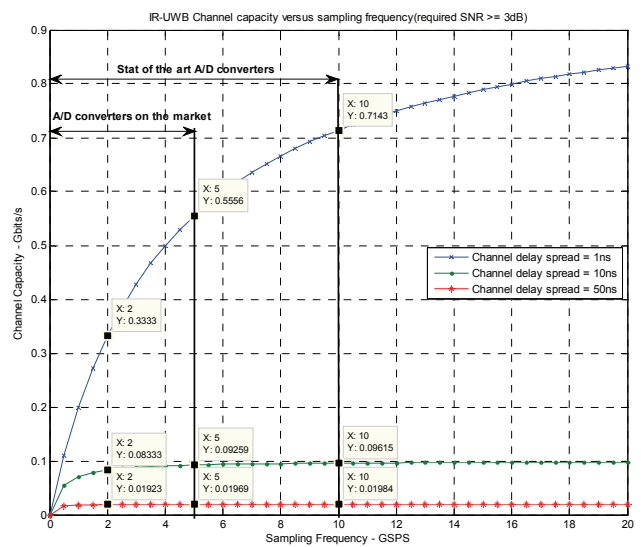

Fig.4 channel Capacity versus sampling frequency regarding to channel delay spread and state of the art for data converters

TABLE I
ACHIEVABLE DATA RATE VALUES FOR MOSTLY DIGITAL IMPLEMENTATIONS, IN FUNCUTION OF REALISTIC CHANNEL DELAY SPREADS AND DATA CONVERTERS PERFORMANCES.

| Environement | RMS delay spread (ns) | Sampling Freq. of A/D (GSPS) | Sampling Factor | Channel Capacity (Mbits/s) |
|---|---|---|---|---|
| Residential LOS | 17 | 2,00 | 4,00 | 52,63157895 |
| Residential LOS | 17 | 5,00 | 4,00 | 56,17977528 |
| Residential LOS | 17 | 10,00 | 4,00 | 57,47126437 |
| Industrial LOS | 9 | 2,00 | 4,00 | 90,90909091 |
| Industrial LOS | 9 | 5,00 | 4,00 | 102,0408163 |
| Industrial LOS | 9 | 10,00 | 4,00 | 106,3829787 |
| Industrial NLOS | 89 | 2,00 | 4,00 | 10,98901099 |
| Industrial NLOS | 89 | 5,00 | 4,00 | 11,13585746 |
| Industrial NLOS | 89 | 10,00 | 4,00 | 11,18568233 |

For example for the industrial LOS channel a data rate of 90 Mbits/s can be attained by using a 2GSPS ADC converter (available on market) with a sampling factor of 4. The no ISI assumption and binary modulations are again considered.

Thanks to table IV and fig. 6, we can see that the data converter performance ($F_s$) is important, as a dimensioning factor, only for low channel delay spreads, regarding the IR-UWB channel capacity for binary modulations. As long as the channel delay spread is large, whatever the sampling frequency, the data rate can't be increased in a significantly manner (for binary modulations). For small channel delay spread, the data converters performances will limit the channel capacity.

### III. IR-UWB CHANNEL CAPACITY FOR ANALOG IMPLEMENTATION

With IR-UWB transceivers with analog implementations, data converters performances are less preponderant than in the mostly digital radio case, since converters are far from the antenna. Direct synthesis no longer exists in analog implementations. Figure 5 exposes one widely use architecture, among analog implementation.

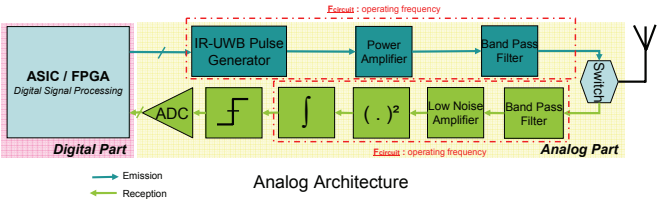

Fig. 5 – Illustration of an IR-UWB analog architecture. The importance of the ADC is reduced in comparison with mostly digital implementation.

With analog implementations, the key parameter is the pulse generator and its ability to generate very short pulse duration at the emitter side. At the receiver side, the limitation is the operating frequency of the circuits. Thus we consider in our expression of the IR-UWB channel capacity for binary modulations (4), only the most constraining frequency, i.e. the minimum one. Note that generally analog operating frequencies are drastically greater than sampling frequencies of data converters.

$$C_{IR-UWB\ ANALOG}(bits/s) = \frac{1}{\frac{1}{F_{circuit}} + d_{RMS}} \quad (4)$$

Where $F_{circuit}$ is the minimal operating frequency of the transceiver among all analog circuits at the emitter and the receiver side; and $d_{RMS}$ is the RMS channel delay spread of the channel.

Figure 6 illustrates the IR-UWB capacity for binary modulation and analog implementation in function of the operating frequency of the circuit. Figure 6 considers three channel delay spreads: 1; 5; and 10 ns.

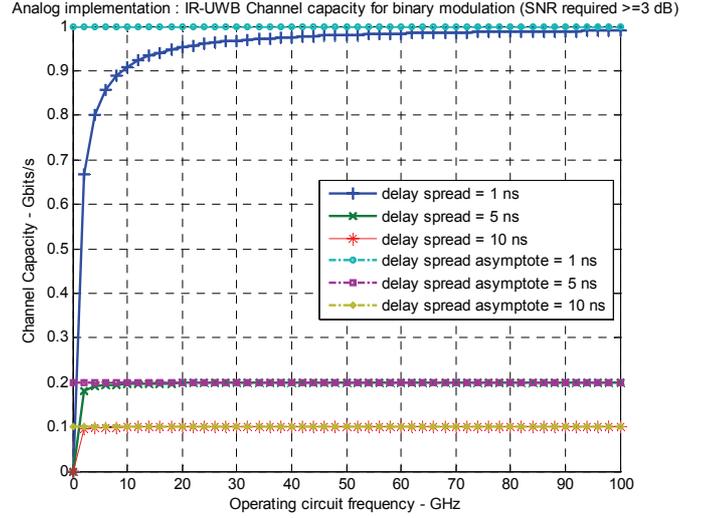

Fig. 6 – IR-UWB channel capacity in the case of binary modulations and analog implementations.

Fig. 8 shows that the maximum IR-UWB channel capacity (the $1/d_{RMS}$ asymptote) is reached quickly when $F_{circuit}$ increase. This fact arrives more quickly in the case of high channel delay spreads than in low channel delay spreads. Typically, in a 10 ns channel delay and for an operating frequency of 5 GHz will give the same data rate capability than a 60GHz operating frequency. Thus, as in the mostly digital case, in the analog implementation the channel delay spread is the preponderant dimensioning parameter. In the case of binary modulations and mono-band schemes, because of the relatively high channel delay spread in UWB realistic scenarios, the increase of the operating frequency is useless.

Thanks to the analysis of the IR-UWB channel capacity, from a high data rate point of view, using an analog implementation is interesting since it could permit to easier achieve high data rate, than in a mostly digital case, by using M-ary modulations. In mostly digital case, M-ary modulations are very difficult due to the antagonism between high sampling frequency data converters and high bit precision data converters. In addition in the analog implementation case, since the delay spread asymptote is attained very quickly, directive antennas can be used for reducing the delay spread of the channel (cf. Table III) and thus profiting in a yielder manner of the high available operating frequency. Directive antennas allow using very high operating frequencies, since the delay spread of the channel is reduced. Furthermore, the higher the working frequency is, the easier directive antennas can be implemented. Thus there is an accenting phenomenon when the operating frequency increases.

As a result with analog techniques, thanks to higher achievable operating frequencies than in the mostly digital case, the channel capacity, in analog cases, can be higher than in the mostly digital case.

Table II exposes some state-of-art IR-UWB pulse generators. Table III summarizes realistic IR-UWB channel delay spreads at 3-10 GHz and 60 GHz, for isotropic and directive antennas, extracted from the IEEE 802.15.4a and the

802.15.3c channel model. With these two kinds of information, we could determine some achievable data rates for analog implementation transceivers. They are listed in table IV, binary and M-ary modulations are also considered.

TABLE II
A SURVEY OF UWB PULSE GENERATOR CAPABILITIES.

| Year | Author | Technology | Pulse duration (ps) | | Ref |
|------|--------|------------|-----|-----|-----|
| | | | min | max | |
| 2007 | Deparis et al. | pHEMT | 50 | 800 | [13] |
| 2007 | Badalawa et al. | CMOS 90 nm | 224 | - | [14] |
| 2006 | Kim et al. | CMOS | 380 | 4000 | [15] |
| 2006 | Bachelet et al. | CMOS 130 nm | 92 | - | [16] |

TABLE III
RMS DELAY SPREAD FOR UWB CHANNEL AT 3-10 AND 60 GHZ IN FUNCTION OF ANTENNAS CONFIGURATIONS

| Residential LOS Channel | Half Power Beam Width | | RMS Delay spread (ns) |
|---|---|---|---|
| | Tx (°) | Rx (°) | |
| UWB 3-10 GHz | 360 | 360 | 17 |
| UWB @ 60GHz | 360 | 360 | 7,718 |
| UWB @ 60GHz | 360 | 60 | 6,2 |
| UWB @ 60GHz | 360 | 15 | 3,455 |
| UWB @ 60GHz | 60 | 60 | 2,147 |
| UWB @ 60GHz | 60 | 15 | 0,948 |
| UWB @ 60GHz | 15 | 15 | 0,87 |

TABLE IV
SOME ACHIEVABLES VALUES OF DATA RATES, FOR IR-UWB ANALOG IMPLEMENTATIONS. IMPACT OF THE CHANNEL DELAY SPREAD AND PULSE GENERATOR CAPABILITIES ARE EXPOSED

| | RMS Delay spread (ns) | Pulse Generator | | Channel capacity (Mbits/s) | | |
|---|---|---|---|---|---|---|
| | | Ref. | Bandwidth (GHz) | Binary Modulations | Ternary Modulations | M=4 Modulations |
| UWB | 17 | [15] | 2,63 | 57,54 | 115,07 | 172,61 |
| | 17 | [14] | 4,46 | 58,06 | 116,12 | 174,17 |
| | 17 | [16] | 10,87 | 58,51 | 117,01 | 175,52 |
| UWB 60 GHz | 7,718 | [16] | 10,87 | 128,04 | 256,08 | 384,12 |
| | 6,2 | [16] | 10,87 | 158,93 | 317,86 | 476,80 |
| | 3,455 | [16] | 10,87 | 281,93 | 563,86 | 845,79 |
| | 2,147 | [16] | 10,87 | 446,63 | 893,26 | 1339,89 |
| | 0,948 | [16] | 10,87 | 961,54 | 1923,08 | 2884,63 |
| | 0,87 | [16] | 10,87 | 1039,51 | 2079,01 | 3118,52 |
| | 0,87 | [13] | 20,00 | 1086,96 | 2173,91 | 3260,87 |

Table IV proves that the most important parameter is the delay spread of the channel, while the operating frequency is a second plan parameter.

## IV. CONCLUSION

Analog and mostly digital implementations impacts on the IR-UWB channel capacity are considered. Whatever the implementation, the channel delay spread is the main limitation. The channel capacity is bounded by a $1/d_{RMS}$ asymptote.

Concerning mostly digital radio, the sampling frequency of data converters, as architecture key point, is used for evaluating the IR-UWB channel capacity. For high channel delay spreads the capacity is limited by the delay spread asymptote. I.e. a sampling frequency change doesn't impact significantly the achievable data rate. Whereas for low channel delay spread the sampling frequency impact the capacity in a direct manner.

The same analysis concerning the delay spread asymptote and the importance of the operating frequency is done for analog implementations. However, in the analog implementation case, operating frequency values are severely larger than state-of-the-art sampling frequencies of data converters. Due to this fact the channel delay spread limitation is achieved more quickly than in mostly digital case. As a result in analog configurations, the channel capacity is almost totally dependent in channel delay spread. Increase the operating frequency is useless from a high data rate point of view. That's why we have exposed in this analog case only, the use of M-ary modulations and directive antenna for achieve higher capacity. Directive antennas reduce the channel delay spread.

At last, for a high data rate criteria comparison, analog solution is more suited for three reasons. M-ary modulations are not viable in mostly digital radio due to ADC performances. The gain of directive antennas is only interesting in analog case. And the sampling frequency is drastically smaller than analog operating frequency due to the $n_{sampling}$ factor (Shannon theorem).

REFERENCES

[1] J. G. Proakis, *"Digital Communications"*, Fourth Edition, McGRAW-HILL, 2001.
[2] Ian O'Donnell, et al., *"An Integrated, Low-Power, Ultra-Wideband Transceiver Architecture for Low-Rate Indoor Wireless Systems"*, Berkeley Wireless Research Center, Univ. of California, Berkeley, IEEE CAS Workshop on Wireless Communications and Networking, Pasadena, 2002.
[3] R. Blaquez et al. *"A Baseband Processor for Impulse Ultra-Wideband Communications"*, IEEE Journal of solid-state circuits, vol. 40, n°9, September 2005.
[4] S. Zhang, et al. *"An MBOK-UWB SoC Transceiver in 0.18 µm CMOS Technology"*, IEEE ASICON 2007, International Conference on ASIC, pp 830-833.
[5] E. R. Green and S. Roy, *"System Architectures for High-rate Ultra-wideband Communication System: A Review of Recents Developments"*, Intel Labs, 2004.
[6] P. Newaskar, *"High speed Data conversion for Digital Ultra-Wideband Radio Receivers"*, Dissertation of a Master of Science in Electrical engineering and Computer Science. Massachusetts Institute of Technology, June 2003.
[7] R. H. Walden, *"Analog-to-digital Converter Survey and Analysis"*, IEEE Journal on Selected Areas in Communications, vol. 17, n°4, April 1999.
[8] W. Yang et al., *"A 3-V 340-mW 14-b 75-Msample/s CMOS ADC with 85-dB SFDR at Nyquist input"*, IEEE Journal of Solid-State Circuits, December 2001.
[9] Y. Akazawa et al*., "A 400 MSPS 8b flash AD conversion LSI"*, IEEE ISSCC 1987, pp 98-99, IEEE February 1987.
[10] I. Mehr et al., *"A 500 Msample/s, 6 bit Nyquist Rate ADC for Disk-Drive Read-Channel applications"*, IEEE Journal of Solid-State Circuits, July 1999.
[11] T. Wakimoto, *"Si Bipolar 2GS/s 6b flash A/D conversion LSI"*, IEEE ISSCC 1988, pp. 232-233.
[12] S. Park et al., *"A 4GS/s 4b Flash ADC in 0.18µm CMOS"*, International Solid-State Circuits Conference, ISSCC 2006, session 31, Very high-speed ADCs and DACs, IEEE 2006.
[13] N. Deparis et al., *"Module radio millimétrique utilisant la synchronisation d'une source 30 GHz par des trains d'impulsions ULB en bande de base"* , Journée Nationale Mircro ondes, JNM 2007, France.
[14] Badalawa et al., *"60 GHz CMOS Pulse Generator",* Electronics Letters, 18[th] January 2007, Vol. 43, n° 2.
[15] Kim et al., *"A tunable CMOS UWB pulse generator"*, ICUWB 2006, International Conference on UWB, pp. 109-112, September 2006.
[16] Bachelet et al., *"Fully integrated CMOS UWB pulse generator"*, Electronics Letters, 26[th] October 2006, Vol. 42, n° 22.